\documentclass[letterpaper]{article}
\usepackage{cite}
\usepackage{amsmath,amsfonts}
\usepackage{algorithm}
\usepackage{algorithmic}
\usepackage{array}
\usepackage[caption=false,font=normal]{subfig}
\usepackage{textcomp}
\usepackage{stfloats}
\usepackage{url}
\usepackage{verbatim}
\usepackage{graphicx}
\def\BibTeX{{\rm B\kern-.05em{\sc i\kern-.025em b}\kern-.08em
    T\kern-.1667em\lower.7ex\hbox{E}\kern-.125emX}}
\usepackage{balance}

\usepackage[rgb]{xcolor}
\usepackage{caption}
\usepackage{array}
\usepackage{enumitem}
\usepackage{hyperref}

\newcolumntype{L}[1]{>{\raggedright\arraybackslash}p{#1}}

\begin{document}

\title{Scalable Multi-Controller Coordination in Periplus via Border-Switch Forwarding Graphs}

\author{E. M. Castro Barbero\textsuperscript{\rm 1}, P. de las Heras Quirós\textsuperscript{\rm 1},\\ F. J. Simó Reigadas\textsuperscript{\rm 1}
\\
\textsuperscript{\rm 1}Universidad Rey Juan Carlos}

\date{July 2026}

\maketitle

\begin{abstract}
In-band SDN control planes, where control traffic shares the data-plane
infrastructure, suit wide-area, resource-constrained deployments---such
as rural backbones---that cannot afford a dedicated control network.
Partitioning such a network across multiple controllers improves
scalability but raises a coordination challenge that in-band designs have
largely ignored: controllers must discover one another and exchange state
in-band, and switches must recover when their controller fails, all
without forwarding state that grows with the number of controllers.

This paper presents the multi-controller coordination plane of Periplus,
an in-band control plane whose single-controller design is developed in a
companion paper. Periplus controllers discover their neighbors through
Controller Advertisement (C-Adv) messages and build inter-controller
routes incrementally: each border switch inserts a partial forwarding
graph covering only the next domain, so per-controller forwarding state
is confined to border switches and never distributed across the interior
of an intermediate domain. The same C-Adv mechanism reattaches a switch
to a surviving controller after a controller failure.

We evaluate a Ryu-based implementation in Mininet, including a 96-switch,
5-controller scenario. Per-switch flow-table state is set by a switch's
role rather than by network size---interior occupancy stays constant as
controllers are added---partitioning scales bootstrap to networks of
around a hundred switches, and inter-controller discovery converges
within seconds. The design needs no switch-firmware modifications: it
runs on stock Open vSwitch, using only its built-in Nicira extensions for
Network Service Header (NSH) encapsulation.

\end{abstract}

\section{Introduction}
\label{sec:introduction}

Despite significant advances in telecommunications infrastructure, many
rural and underserved regions still lack reliable Internet
connectivity~\cite{lorenti2024rural}. These regions combine low
population density, limited income, a shortage of local technical
personnel, and scarce complementary power and communications
infrastructure.
Their service requirements, however, are often similar to those of dense
urban areas, so they call for technologies that are simultaneously
low-cost, robust, flexible, and easy to deploy and operate.

Software-defined networking (SDN)~\cite{kreutz2014software} can bring
clear benefits to these networks, but only if some of their specific
constraints are met. Because they lack a parallel infrastructure for
control or signaling, the control plane must run in-band, sharing the
data-plane links it manages~\cite{jalili2017comprehensive, surveyUAH}.
SDN's centralized control simplifies operation, yet it must be reinforced
with robustness and resilience mechanisms so that the network keeps
serving users despite the unstable links and frequent topology changes
typical of these environments, and continues to operate with little to no
human intervention.

An in-band control plane must overcome four challenges that a dedicated
control network would not face. First, a switch must attach
to the control plane before any forwarding rule exists to carry its own
control traffic (\emph{bootstrapping}). Second, once attached, its control
messages must travel over efficient paths that keep per-switch state low
(\emph{routing}). Third, switches must survive link and node failures
without waiting for the controller to intervene (\emph{failure recovery}).
Fourth, once the network is large enough to require several controllers,
those controllers must stay coordinated over the same in-band substrate
they manage (\emph{multi-controller coordination}).

Earlier proposals for these environments strike different balances: some
prioritize adaptability, low cost, and quality at the expense of high
complexity~\cite{tucan3g}, while others trade some quality for easier
deployment and operation~\cite{wiback}. Periplus was recently proposed as
an SDN architecture tailored to these settings. Its single-controller
design, presented in~\cite{periplus-single}, addresses the first three
challenges: a bootstrap that installs flow rules in only two switches when
a switch joins, graph-based source routing with minimal per-switch state,
and sub-50\,ms failure recovery performed locally at the switch, combining
prompt link-failure detection with forwarding alternatives precomputed for
each hop. That design, however, assumes a single controller and does not
address the fourth challenge. The present paper builds on Periplus to
tackle this fourth and least-explored challenge: coordinating multiple
controllers over the in-band network.

Partitioning a wide-area network across several controllers improves
scalability and keeps control traffic local, but coordinating those
controllers in-band is difficult. Controllers must discover one another
and exchange switch state over the same network they manage, and a
switch must keep reaching a controller when its own fails, all without
flooding the network or storing forwarding state that grows with the
number of controllers. Existing in-band designs that admit multiple
controllers (Section~\ref{sec:related-work}) give each controller its
own spanning tree or a set of disjoint control paths, so per-switch
forwarding state scales with the controller count, eroding the
scalability that motivated partitioning in the first place.

This paper contributes a multi-controller coordination mechanism whose
per-controller forwarding state is confined to border switches.
Periplus controllers discover their neighbors through Controller
Advertisement (C-Adv) messages and establish inter-controller routes
across domains incrementally: each border switch inserts a partial
forwarding graph covering only the next domain, so the routing state
needed to reach a remote controller is never distributed across the
interior switches of any intermediate domain. The same C-Adv substrate
lets a switch recover after a controller failure by re-attaching
through a neighboring domain. Because coordination state is confined
to domain boundaries, the architecture remains scalable as the number
of controllers and switches increases.

Periplus is implemented in Python on the Ryu SDN
framework~\cite{ryuDoc, ryuBook} and evaluated through network emulation
in Mininet across several multi-controller topologies, including a
large-scale scenario with 96 switches and 5 controllers. The results
confirm that per-switch forwarding state stays bounded: interior switches
hold a constant number of rules regardless of the number of controllers,
and additional state appears only at border switches.

The remainder of this paper is organized as follows.
Section~\ref{sec:related-work} reviews related work on multi-controller
coordination in in-band SDN. Section~\ref{sec:sdn-control-plane} recaps
the single-controller design of Periplus that the multi-controller
plane builds on. Section~\ref{sec:multiple-controllers} presents the
multi-controller coordination mechanism, covering inter-controller
discovery, communication, and switch recovery after a controller
failure. Section~\ref{sec:evaluation} reports an experimental
evaluation in Mininet across multi-controller topologies, including the
96-switch, 5-controller scenario. Finally,
Section~\ref{sec:conclusions} concludes the paper.

\section{Related Work}
\label{sec:related-work}

This paper extends Periplus, an in-band SDN control plane, to operate
across multiple controllers. The single-controller design of Periplus
(automatic bootstrapping, graph-based source routing, and sub-50\,ms
failure recovery) and the related work on those three challenges are
presented in~\cite{periplus-single}. This section
concentrates on the fourth and least-explored challenge of in-band SDN
control planes: coordination among multiple controllers.

For context, we briefly situate the three single-controller
challenges. Bootstrapping proposals such as Medieval and
Renaissance~\cite{schiff2016ground,canini2022renaissance}, Sakic et
al.~\cite{sakic2020automated}, FASIC~\cite{su2017fasic}, and Wong and
Lee~\cite{wong2023design} achieve plug-and-play attachment but at the
cost of network-wide flooding or per-switch state that grows with
network size. For routing, source-based approaches such as
Amaru~\cite{lopez2019amaru} reduce forwarding state yet cannot
guarantee controller reachability after a failure. For failure
recovery, protection schemes such as those of Sharma et
al.~\cite{sharma2013automatic,sharma2013fast,sharma2016band} reach
sub-50\,ms recovery but preinstall backup paths on every switch and
rely on STP for bootstrap.

\textbf{Multi-controller coordination.} Multi-controller coordination
has received comparatively little attention in the in-band control
plane literature. Medieval and Renaissance support multiple
controllers by having each controller build an additional spanning
tree covering all switches in the network, which imposes
per-controller forwarding state on every switch. Sakic et
al.~\cite{sakic2020automated} deploy multiple controller replicas
synchronized via RAFT consensus, provisioning disjoint control paths
per switch-controller pair; per-switch forwarding state likewise grows
with the number of controllers. Holzmann et
al.~\cite{holzmann2018towards} extend Izzy to elastic controller
clusters via the Seedling algorithm, which divides controllers into
proximity-based groups and assigns a separate Izzy spanning tree per
group; however, per-switch forwarding state still scales with the
number of controllers. Chan et al.~\cite{chan2018fast} target fast
failure recovery in in-band multi-controller OpenFlow networks: a Main
Controller and several Standby Controllers collaborate to detect
failures through monitoring cycles and reroute traffic by installing
new flow entries, provisioning $K$ disjoint control paths per switch
(one from each controller) to guarantee reachability after any single
device failure, with average recovery times below 50\,ms. Unlike
Periplus, recovery there requires explicit coordination between
controllers and is not performed locally at the switch. Across these
proposals, none jointly addresses fast inter-controller communication
and bounded per-switch state overhead in an in-band setting.

Schiff et al.~\cite{schiff2016band} explore an orthogonal approach to
controller coordination: using part of the data-plane configuration
space as transactional shared memory, implementing atomic
compare-and-swap operations via OpenFlow bundling to synchronize
distributed controllers.

Panda et al.~\cite{panda2017scl} argue that strong consensus protocols
are conceptually inappropriate for network state in distributed SDN
controllers, since the physical network is inherently uncertain; they
propose eventual correctness instead. Periplus is aligned with this
view: it provides the inter-controller communication infrastructure
without imposing any consistency model, leaving coordination semantics
to the applications above.

Multi-controller coordination thus remains the least explored
challenge of in-band SDN control planes, and existing solutions impose
forwarding state proportional to the number of controllers on every
switch. Periplus departs from this pattern: it uses Controller
Advertisement messages and border switches to confine per-controller
forwarding state to the switches at domain boundaries, and it provides
the inter-controller communication infrastructure without mandating
any consistency model. The following sections describe this mechanism
and evaluate how it scales Periplus across multiple controllers.

\section{Background: Single-Controller Periplus}
\label{sec:sdn-control-plane}

This section recaps the single-controller design of Periplus to the
extent needed to follow its multi-controller extension; the full
design, with its protocols, tables, and message-overhead analysis, is
presented in~\cite{periplus-single}. Periplus
rests on three core design choices for the single-controller case:
anchor-relayed bootstrap, graph-based source routing, and
packet-encoded failure recovery. A fourth mechanism, Controller
Advertisement (C-Adv) messages, serves as a cross-cutting discovery
substrate. Multi-controller operation
(Section~\ref{sec:multiple-controllers}) extends these choices through
C-Adv messages and border switches.

We assume that node communication interfaces, wired or wireless, are
based on IEEE 802 link technologies, so that each interface is
identified by a unique 48-bit MAC address, and that links are
bidirectional. Switches are denoted $si$ with $cj$ reserved for the
OVS switch co-located with the controller (the root switch).

\subsection{Bootstrap}
\label{sec:bootstrap}

Periplus bootstraps the control plane incrementally: switches attach
one by one, each relying on a neighboring managed switch to relay its
connection requests. Before it can connect, an unmanaged switch must
learn the port toward the controller, which it caches in a soft-state
register ($reg9$) with a short hard timeout. It learns this port from
whichever controller-attributable signal arrives first: a proxy-ARP
reply from a managed neighbor (the anchor) answering on the
controller's behalf, or a Controller Advertisement re-forwarded by a
managed neighbor (Section~\ref{sec:c-adv}). Both can only reach an
unmanaged switch from the controller side, so the ingress port on which
they arrive is the upstream port. The two triggers cover complementary
cases: the ARP reply handles a cold start, and the C-Adv handles a warm
restart, when a previously managed switch still caches the controller's
MAC address, issues no ARP request, and can relearn its upstream port
only from a neighbor's C-Adv. The switch sends its TCP SYN through the
learned port; if none has been learned the SYN is dropped, and if the
signals stop the cached entry expires and the switch reverts to the
unmanaged state. The controller identifies the new switch from the
Ethernet source address of its SYN and installs the forwarding flows
that place it in the managed state, updating only two switches rather
than every switch on the path to the controller; the full preconfigured
rule set is given in~\cite{periplus-single}. In a multi-controller
deployment this anchor relay operates within each controller domain, and
border switches extend it across domain boundaries
(Section~\ref{sec:multiple-controllers}).

\subsection{Forwarding based on graphs}
\label{sec:forwarding-graphs}

Periplus routes control traffic with the Slick Packets
approach~\cite{nguyen2011slick}: a forwarding graph, an acyclic
subgraph of the topology embedded in the packet headers between the L2
and L3 layers using NSH~\cite{rfc8300} (realized with OVS's Nicira
extensions~\cite{Nicira}), encodes a primary path $P$
and, at each hop $i$, an ordered list of up to $K_i$ alternative paths
as output-port sequences. The controller computes $P$ with Dijkstra
and derives source-disjoint alternatives by iteratively pruning and
re-running it. A switch that cannot forward through its primary port
falls over locally to the first reachable alternative, with no
controller involvement. Only the root switch co-located with the
controller maintains routing state for all switches in its domain;
every other switch stores only the path to its own controller. This
domain-local property is what the multi-controller plane builds on:
inter-domain routes are assembled from partial graphs inserted by
border switches, so no interior switch ever holds routing state for a
remote controller (Section~\ref{sec:multiple-controllers}).

\subsection{Monitoring ports}
\label{sec:monitoringports}

Each switch monitors its links with two complementary detectors:
Bidirectional Forwarding Detection (BFD)~\cite{RFC5880}, used together
with OpenFlow select group tables~\cite{specification2015open}, and
Input Traffic Detection (ITD),
based on OpenFlow Echo Request/Reply messages and learning flows. On
detecting a link or node failure, a switch immediately reroutes along
an encoded alternative from the packet's graph, meeting carrier-grade
(sub-50\,ms) recovery without contacting the controller.

\subsection{Link Discovery Protocol}
\label{sec:c-adv}

Once switches are managed, the controller knows the tree of links that
connect it to them, but not the redundant links that can serve as
alternatives. Periplus discovers these through Controller
Advertisement (C-Adv) messages. The controller periodically issues a
single Packet-Out to the root switch, which floods it; each switch
floods a C-Adv only on first reception, suppressing duplicates with a
short-lived learning flow whose hard timeout is shorter than the C-Adv
period, and reports the reception to the controller with a Packet-In.
Each managed switch encodes its own identity and egress port into the
header of every C-Adv it forwards, so the controller can identify both
endpoints of each discovered link. This controlled flooding avoids
broadcast storms and reduces the controller to a single Packet-Out per
discovery round; its message-overhead advantage over standard
link-discovery protocols is analyzed in~\cite{periplus-single}. C-Adv
is the substrate the multi-controller plane reuses:
it underpins both inter-controller discovery and the recovery of
switches after a controller failure
(Section~\ref{sec:multiple-controllers}).

\section{Multiple Controllers}
\label{sec:multiple-controllers}

Periplus is designed to support multiple controller instances. Each
controller manages a disjoint set of switches, known as its domain; a
switch with a link into another controller's domain is a \emph{border
switch}, through which inter-controller traffic transits from one domain
to the next. A switch is assigned to a controller as a direct
consequence of the
bootstrap mechanism of Section~\ref{sec:bootstrap}: the nearest
managed neighbor relays the booting switch's TCP SYN toward its own
controller, yielding a first-come, first-served (FCFS) assignment
to typically the nearest controller. Disjointness is enforced implicitly,
since once the routing flows are installed the switch's control
traffic is directed exclusively through its anchor to a single
controller; if the switch later reboots, it re-runs the bootstrap
procedure and may attach to a different controller if the topology
has changed.

This assignment rests on how controllers are addressed. All controllers
share a single anycast address, so an unmanaged switch attaches to
whichever controller its bootstrap signals reach first---typically the
nearest---without needing to know which controller that is. In addition,
each controller owns a distinct unicast address that identifies it to
the other controllers; the inter-controller discovery and communication
mechanisms described below rely on these addresses to tell controllers
apart and route between them. These unicast addresses are used only
among controllers: a switch never needs to know the unicast address of
any controller, and reaches the control plane solely through the shared
anycast address.

This section describes how controllers discover and
communicate with neighbor controllers, how this mechanism is extended
to non-neighbor controllers, and how switches recover from a
controller failure.

\subsection{Communication between neighbor controllers}
\label{sec:icd-neighbor}

C-Adv messages are used to discover other controllers. The propagation
of C-Adv messages follows the controlled flooding mechanism described
in Section~\ref{sec:c-adv}. Since multiple controllers coexist within
the same network, when a switch forwards a C-Adv through an interface
connected to a switch managed by another controller, the receiving
switch generates a Packet-In message to its own controller and does
not forward the C-Adv further. This mechanism enables a controller to
detect the presence of another controller in the network.

\begin{figure}[!h]
    \centering

    \subfloat[s1 receives C-Adv from controller~2.\label{fig:c-adv-2c-1}]
    {
        \includegraphics[width=0.9\linewidth]{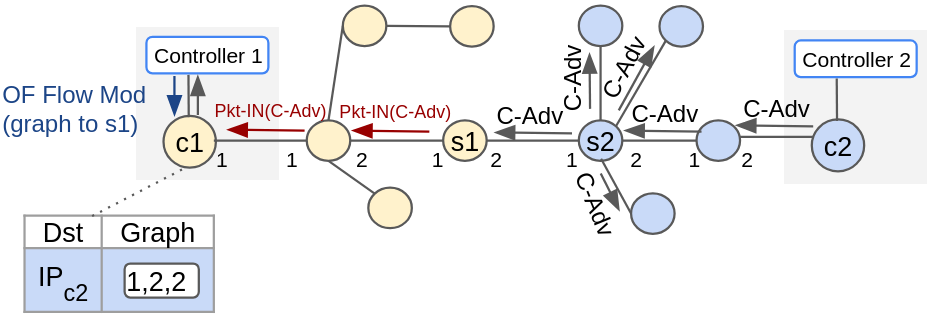}
    }


    \subfloat[controller~1 sends a message to controller~2.\label{fig:c-adv-2c-2}]
    {
        \includegraphics[width=0.9\linewidth]{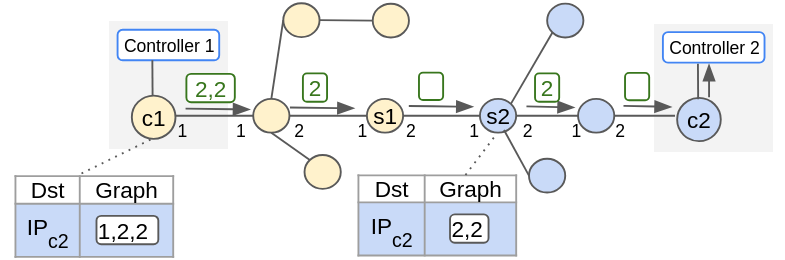}
    }

    \caption{Inter-controller discovery via C-Adv on a two-controller topology: (a) $s1$ receives a C-Adv from controller~2 and reports the link to controller~1; (b) controller~1 forwards a packet to controller~2 using the embedded forwarding graph.}
    \label{fig:c-adv-2c}
\end{figure}

Figure~\ref{fig:c-adv-2c} illustrates a network with two
controllers. Switches managed by controller~1 are shown in yellow,
while those managed by controller~2 are shown in blue. As depicted in
Fig.~\ref{fig:c-adv-2c-1}, switch $s1$ receives a C-Adv message
generated by controller~2 and generates a Packet-In to
controller~1. Upon receiving this message, controller~1 becomes aware
of controller~2's presence, and then controller~1 installs a new flow
in $c1$ (the root switch of controller~1) to reach controller~2 via
$s1$. This flow instructs $c1$ to embed in packets destined for $c2$ a
forwarding graph whose primary path terminates at port~2 of $s1$.  Figure~\ref{fig:c-adv-2c-2}
shows how packets from controller~1 to controller~2 are forwarded
through controller~1's domain using the embedded graph until reaching
$s1$. Between $s1$ and $s2$, packets do not carry the graph; however,
$s2$ is a border switch with a flow installed to embed the forwarding
graph for its own domain, and therefore inserts the graph that enables
the packet to reach controller~2.

Analogously, when $s2$ receives a C-Adv from controller~1 and
generates a Packet-In to its own controller, controller~2 installs a
new flow rule in $c2$ to reach controller~1.

\subsection{Communication between non-neighbor controllers}
\label{sec:icd-nonNeighbor}

The mechanism described above operates directly when controller
domains are neighbors. It generalizes to non-neighboring controllers
through Algorithm~\ref{alg:controller_discovery}.

\begin{algorithm}
\caption{Inter-Controller Discovery (at controller $c$)}
\label{alg:controller_discovery}
\begin{algorithmic}[1]
\small\raggedright
\medskip
\STATE \textbf{State:} $\mathit{known}(c)$: set of entries $(c_x,\ \mathit{is\_neighbor},\ \mathit{learnt\_from},\ \mathit{alive})$, where:
\STATE \hspace{2em} $\mathit{is\_neighbor}$: \textbf{true} if $c_x$ discovered directly via Packet-In
\STATE \hspace{2em} $\mathit{learnt\_from}$: path vector (controllers that reported $c_x$)
\STATE \hspace{2em} $\mathit{alive}$: liveness counter, reset to $T$ on each C-Adv refresh
\medskip
\STATE \textbf{upon} \textit{C-Adv from $c_n$ carrying $P = \{(c_x,\ \mathit{learnt\_from}_x)\}$} \textbf{do}
\STATE \hspace{1.5em} add or refresh $c_n$ in $\mathit{known}(c)$:
\STATE \hspace{3em} $\mathit{is\_neighbor} \leftarrow \textbf{true}$,\ $\mathit{learnt\_from} \leftarrow [c_n]$,\ $\mathit{alive} \leftarrow T$
\STATE \hspace{1.5em} update ARP cache for $c_n$
\STATE \hspace{1.5em} install forwarding graph to reach $c_n$
\STATE \hspace{1.5em} \textbf{for each} $(c_x,\ \mathit{learnt\_from}_x) \in P$ \textbf{do}
\STATE \hfill $\triangleright$ \textit{skip ourselves}
\STATE \hspace{3em} \textbf{if} $c_x = c$: \textbf{continue}
\STATE \hfill $\triangleright$ \textit{neighbor takes precedence}
\STATE \hspace{3em} \textbf{if} $\mathit{is\_neighbor}_x = \textbf{true}$: \textbf{continue}
\STATE \hfill $\triangleright$ \textit{loop detected}
\STATE \hspace{3em} \textbf{if} $c \in \mathit{learnt\_from}_x$: \textbf{continue}
\STATE \hspace{3em} add or refresh $c_x$ in $\mathit{known}(c)$:
\STATE \hspace{3.5em} $\mathit{is\_neighbor} \leftarrow \textbf{false}$,
\STATE \hspace{5em} $\mathit{learnt\_from} \leftarrow \mathit{learnt\_from}_x + [c_n]$,
\STATE \hspace{5em} $\mathit{alive} \leftarrow T$
\STATE \hspace{3em} update ARP cache for $c_x$
\STATE \hspace{3em} install forwarding graph to reach $c_x$
\medskip
\STATE \textbf{upon} \textit{periodic timer} \textbf{do}
\STATE \hspace{1.5em} \textbf{for each} $c_x \in \mathit{known}(c)$: decrement $\mathit{alive}_x$
\STATE \hspace{1.5em} \textbf{if} $\mathit{alive}_x = 0$: remove $c_x$ from $\mathit{known}(c)$ and its flows
\STATE \hspace{1.5em} refresh forwarding graph to each surviving $c_x \in \mathit{known}(c)$
\STATE \hspace{1.5em} install border routes between all neighbor controller pairs
\STATE \hspace{3em} $\triangleright$ \textit{see state distribution analysis below}
\medskip
\STATE C-Adv sent by $c$ carries $\{(c_x,\ \mathit{learnt\_from}_x)\}$ for all $c_x \in \mathit{known}(c)$
\end{algorithmic}
\end{algorithm}

Each C-Adv carries a list of all controllers known to the sender, each
entry accompanied by its $\mathit{learnt\_from}$ path vector. Upon
reception, the local controller processes the sending controller as a
direct neighbor, then iterates over the payload to discover
non-neighboring controllers. Three conditions guard each entry: the
controller's own IP is skipped; a direct neighbor cannot be downgraded
to non-neighbor status; and any entry whose $\mathit{learnt\_from}$
list already contains the local controller's IP is discarded. This
last condition prevents routing loops by the same principle as the
AS\_PATH attribute in BGP~\cite{rfc4271}: just as a BGP router rejects
a route whose AS\_PATH already contains its own AS number, a
controller rejects an entry that has already passed through it.

Forwarding graphs are computed using Dijkstra's algorithm on the local
topology graph: the primary path is the shortest path within the local
domain to the border switch adjacent to the destination controller's
domain, and for each hop, an alternative path is computed where one
exists. These flows are installed when a new peer controller is
discovered, following the same soft-state principle as the
$\mathit{alive}$ counter.

This locality has a direct consequence for state
distribution. Non-border switches require no state beyond the path to
their own controller---the same property established in
Section~\ref{sec:forwarding-graphs} for the single-controller
case. Additional state is installed only at border switches that serve
as entry points for transit between pairs of neighbor controllers
(Algorithm~\ref{alg:controller_discovery}, line~27): each such switch
stores a single forwarding graph per transit destination, encoding
only the short, local path within its own domain from the entry border
to the exit border switch facing the next domain. This guarantees
transit without distributing state across the domain interior.

Liveness is maintained through a soft-state $\mathit{alive}$ counter,
following the same design principle as RSVP~\cite{rfc2205}: entries
are periodically refreshed by incoming C-Adv messages and expire
silently when refreshes stop. When a controller fails, its C-Adv
messages cease, all entries referencing it are no longer refreshed,
and they eventually expire across the network.

\begin{figure}[ht]
\centering
	\includegraphics[width=\linewidth]{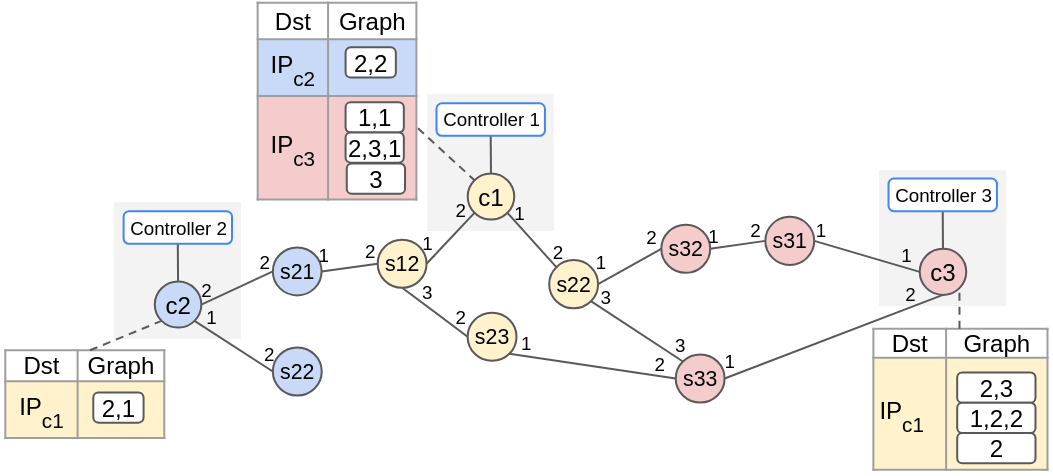}
	\caption{Three-controller topology where controllers~2 and~3 are
non-neighbors and discover each other via controller~1's C-Adv
advertisements.}
\label{fig:neigh_controllers}
\end{figure}

Fig.~\ref{fig:neigh_controllers} illustrates
Algorithm~\ref{alg:controller_discovery} for a three-controller
topology. Controller~1 is a direct neighbor of both controller~2 and
controller~3. It advertises both in its C-Adv. Controllers~2 and~3
learn of each other from controller~1's payload and install the
required flows; the loop-detection condition prevents each from
re-accepting the other's entry via controller~1.

\begin{figure*}[tp]
    \centering

	\subfloat[New flow rule in $c2$ to reach controller~3.\label{fig:no_neigh_controllers_c2}]
    {
        \resizebox{0.5\linewidth}{!}{\includegraphics{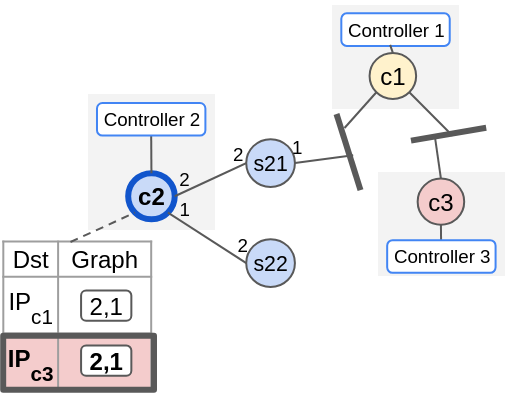}}
    }

    \vspace{1ex}

	\subfloat[New flow rule in $c3$ to reach controller~2.\label{fig:no_neigh_controllers_c3}]
    {
        \resizebox{0.6\linewidth}{!}{\includegraphics{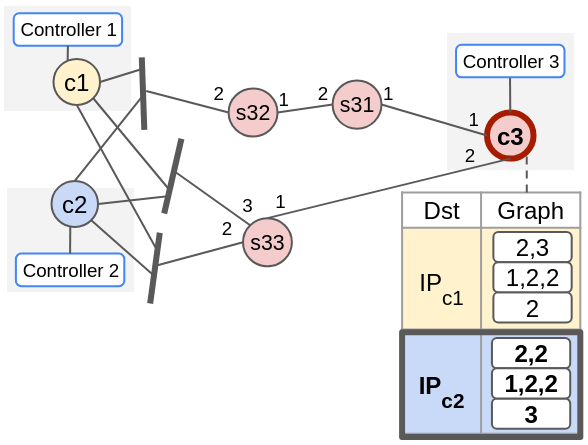}}
    }

    \vspace{1ex}

	\subfloat[New flow rules in border domain switches to reach controllers.\label{fig:no_neigh_controllers_c1}]
    {
        \resizebox{0.6\linewidth}{!}{\includegraphics{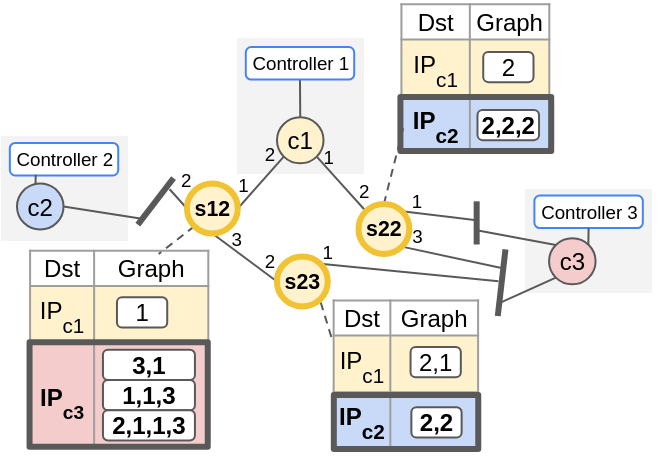}}
    }

    \caption{Non-neighbor controller discovery (same topology as Fig.~\ref{fig:neigh_controllers}); switches in other domains are collapsed into black segments.}
    \label{fig:controller_discovery}
\end{figure*}

Fig.~\ref{fig:controller_discovery} uses the same three-controller
topology as Fig.~\ref{fig:neigh_controllers}, showing the forwarding
state each controller holds after discovery completes; in each
subfigure, switches belonging to other domains are collapsed into
black segments. As shown in Fig.~\ref{fig:no_neigh_controllers_c2} and
Fig.~\ref{fig:no_neigh_controllers_c3}, controllers~2 and~3 each
install a new flow in their root switches ($c2$ and $c3$) to embed the
connectivity graph required to reach the newly discovered
non-neighboring controller. Controller~1 also installs new flows in
its border domain switches: as illustrated in
Fig.~\ref{fig:no_neigh_controllers_c1}, a new flow in switch $s12$ to
reach controller~3, and new flows in switches $s23$ and $s22$ to reach
controller~2.

This mechanism resembles a segment-routing approach~\cite{rfc8402}, but the
forwarding graph is installed incrementally rather than at the
ingress: each border switch adds the segment crossing its own
domain. For example, a message from controller~2 to controller~3 in
the topology of Fig.~\ref{fig:neigh_controllers} carries $P=[2,1]$
from $c2$ to border switch $s12$, then $P=[3,1]$ across
controller~1's domain to the exit border switch $s33$, where the
local segment within controller~3's domain is added. The full
end-to-end path is never encoded at any single point, which scales
more effectively than ingress-only encoding across multiple domains
containing large numbers of switches.

\subsection{Controller Failure Recovery}
\label{sec:cfrProcedure}

Unlike the link and node failures of
Section~\ref{sec:monitoringports}, which a switch survives locally by
rerouting to an alternative path toward the same controller, a
controller failure leaves the switch unable to reach its controller at
all, so it must attach to a surviving one. The affected switches
recover through the event-driven procedure of
Algorithm~\ref{alg:failure_recovery}, which reuses the preconfigured
bootstrap rules of Section~\ref{sec:bootstrap}.

\begin{algorithm}
\caption{Controller Failure Recovery (at switch $s$)}
\label{alg:failure_recovery}
\begin{algorithmic}[1]
\small\raggedright
\medskip
\STATE \textbf{upon} \textit{controller-installed routing flows expiring} \textbf{do}
\STATE \hspace{1.5em} $s$ reverts to preconfigured initial flow rules
\medskip
\STATE \textbf{upon} \textit{C-Adv from neighboring managed switch} \textbf{do}
\STATE \hspace{1.5em} $reg9 \leftarrow$ ingress port of C-Adv
\STATE \hspace{1.5em} send TCP SYN via $reg9$
\medskip
\STATE \textbf{upon} \textit{$reg9 = 0$ and ARP cache empty} \textbf{do}
\STATE \hspace{1.5em} broadcast ARP request through all ports
\STATE \textbf{upon} \textit{ARP reply received} \textbf{do}
\STATE \hspace{1.5em} $reg9 \leftarrow$ ingress port of ARP reply
\STATE \hspace{1.5em} send TCP SYN via $reg9$
\medskip
\STATE \textbf{upon} \textit{TCP session established with new controller} \textbf{do}
\STATE \hspace{1.5em} complete OpenFlow handshake
\STATE \hspace{1.5em} new controller installs routing flows
\STATE \hspace{1.5em} $s$ transitions to managed state under new domain
\end{algorithmic}
\end{algorithm}

The controller installs the routing flows with a hard timeout and
refreshes them periodically. Recovery therefore requires no explicit
failure detection on the switch's part: it is triggered implicitly when
the controller fails and stops refreshing these flows, so they expire and
return the switch to the unmanaged state of Section~\ref{sec:bootstrap}
with no upstream port cached ($reg9 = 0$). The two learning signals that drive
initial attachment then reattach the switch, now to a surviving
controller. As at bootstrap, the C-Adv path takes priority: a Controller
Advertisement from a managed neighbor sets $reg9$ and suppresses the ARP
path, which activates only when no C-Adv has arrived and the ARP cache
holds no controller entry (the warm-restart case of
Section~\ref{sec:bootstrap}). Both paths converge at the TCP SYN,
re-running the attachment of Section~\ref{sec:bootstrap} against the
surviving controller, which installs routing flows and brings the switch
into its domain.

\section{Evaluation}
\label{sec:evaluation}

This section evaluates the multi-controller plane of Periplus along the
two properties its design promises (Section~\ref{sec:introduction}):
that per-controller forwarding state stays confined to border switches,
and that partitioning a network across controllers scales the control
plane to networks of around a hundred switches, with each controller
managing only a small domain. We first measure how bootstrap time scales
as a network is partitioned across an increasing number of controllers
(Section~\ref{sec:bootstrapScaling}). We then evaluate the coordination
mechanism in action: the time for controllers to discover one another
(Section~\ref{sec:interControllerDiscovery}) and the time for switches
to recover after a controller fails
(Section~\ref{sec:controllerFailureRecovery}). Finally, we measure the
per-switch flow-table footprint as a function of each switch's role,
confirming that coordination state is bounded at domain boundaries
(Section~\ref{sec:flowtableoccupancy}).

The experimental setup follows that of~\cite{periplus-single}.
Topologies are emulated with Mininet~2.3~\cite{deOliveira2014mininet},
switches are OVS~2.15, and Periplus runs on the Ryu~4.32 SDN framework,
on an Ubuntu 24.04 system with an AMD Ryzen 7 4800H CPU and 64~GB of
RAM. As in the single-controller evaluation, each OVS instance runs in
its own network namespace and in secure mode
(\texttt{fail-mode=secure}), so that a switch cannot reach any
controller until a neighbouring switch is itself managed, reproducing a
pure in-band control plane. Unless otherwise noted, each experiment is
repeated twenty times.

\begin{figure}[hbt]
\centering
\includegraphics[width=\linewidth]{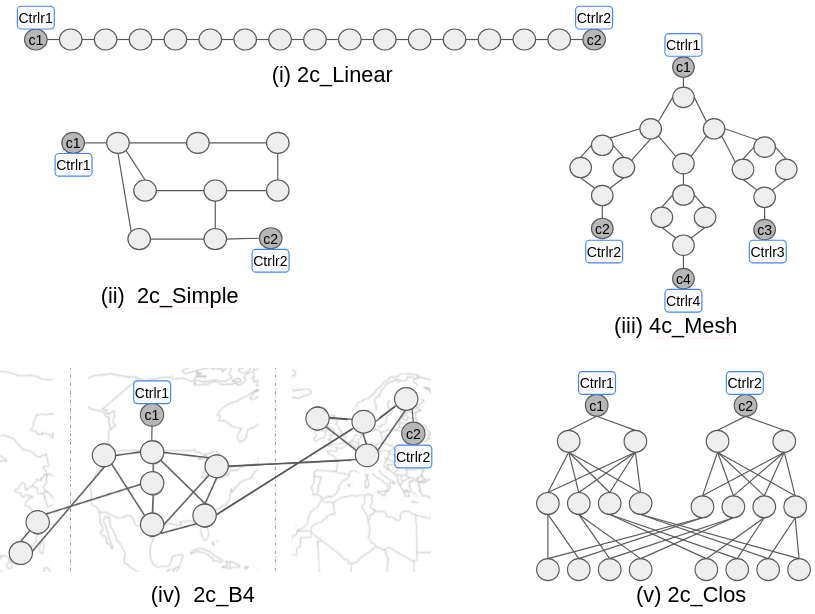}
\caption{Multi-controller topology variants: (i) 2c\_Linear, (ii)
  2c\_Simple, (iii) 4c\_Mesh, (iv) 2c\_B4, and (v) 2c\_Clos. The prefix
  indicates the number of controllers.}
\label{fig:multipleControllers}
\end{figure}

\begin{figure}[hbt]
\centering
\includegraphics[width=0.7\linewidth]{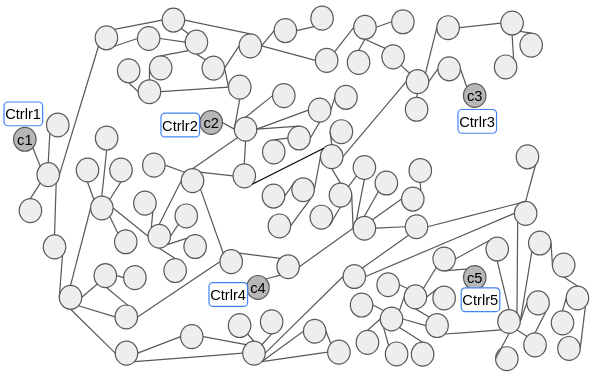}
\caption{Large-scale network topology (Large) with 96 switches and 5
  controllers placed at geographically distributed positions.}
\label{fig:att}
\end{figure}

We evaluate the multi-controller variants of
Fig.~\ref{fig:multipleControllers}, derived from the single-controller
topologies of~\cite{periplus-single} by placing controllers at selected
switches. Two additional Mesh variants (2c\_Mesh and 3c\_Mesh) are
obtained from the 4c\_Mesh topology by removing controllers. The
large-scale topology of Fig.~\ref{fig:att} (Large, 96 switches) is
tested with 2, 3, 4, and 5 controllers.

\subsection{Bootstrap scaling across controllers}
\label{sec:bootstrapScaling}

Periplus bootstraps incrementally, so a switch can begin attaching only
once a neighbouring switch is managed: bootstrap time is governed by the
diameter of the region a controller must reach, not by the total number
of switches~\cite{periplus-single}. Partitioning a network across
several controllers exploits this directly, since each controller
bootstraps only its own domain. Adding controllers shrinks the
per-controller diameter and shortens the critical path, letting the
control plane scale to networks far larger than a single controller
could bootstrap quickly. This scaling is what the bounded per-controller
state of Section~\ref{sec:flowtableoccupancy} pays for: because
coordination state stays at domain boundaries, partitioning adds
controllers without inflating per-switch state.

We measure the wall-clock bootstrap time, the interval from the first
TCP SYN received at any controller to the moment the last switch is
marked managed across all controller logs, on the
Large topology of Fig.~\ref{fig:att} (96 switches) partitioned across 2,
3, 4, and 5 controllers. We use twenty repetitions per configuration on
the patched OVS build (Section~\ref{sec:controllerFailureRecovery}
describes the patch).

\begin{figure}[h]
    \centering
    \includegraphics[width=0.5\linewidth]{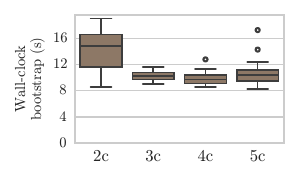}
\caption{Wall-clock bootstrap for the Large topology (96 switches) with
  2, 3, 4, and 5 controllers.}
\label{fig:attScaling}
\end{figure}

Fig.~\ref{fig:attScaling} shows the result. The median wall-clock falls
from $14.8$~s with two controllers to $10.2$~s with three, and then
flattens ($9.7$~s at four, $10.4$~s at five): once the per-controller
diameter reaches its minimum, adding controllers yields no further gain.
A 96-switch network thus bootstraps in around ten seconds once it is
partitioned into three or more domains, confirming that partitioning
keeps each controller's domain small---a dozen switches
here---while the network as a whole scales to nearly a hundred. The
within-configuration spread
reflects the first-come-first-served variability in how many switches
each controller ends up managing (Section~\ref{sec:multiple-controllers}).

\subsection{Inter-controller discovery time}
\label{sec:interControllerDiscovery}

In a multi-controller deployment, no inter-controller coordination can
begin until every controller has discovered every other controller. The
convergence time of this initial discovery phase is therefore the floor
on how quickly the multi-controller plane becomes fully operational. The
C-Adv-based mechanism of Section~\ref{sec:multiple-controllers} performs
this discovery implicitly: each controller learns about a peer the first
time one of its managed switches forwards a C-Adv originated by that
peer. We measure how long that learning takes in practice.

We run three multi-controller scenarios (2c\_Simple, 4c\_Mesh, and the
5-controller large-scale topology), and for each run we record the
interval between the moment the last controller is launched and the
latest ``first-time-learned'' event across all controllers. We use
twenty repetitions per scenario.

\begin{figure}[h]
    \centering
    \includegraphics[width=0.65\linewidth]{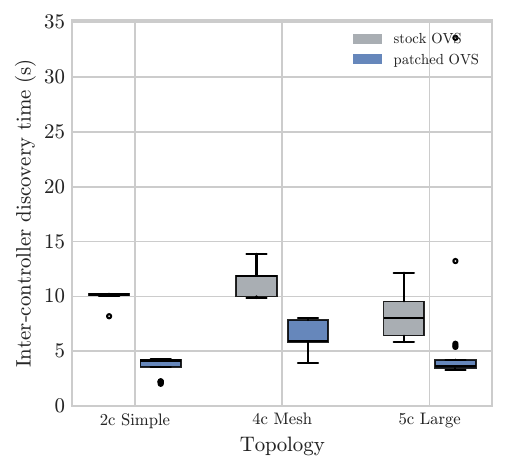}
\caption{Inter-controller discovery time, measured from the launch of
  the last controller until every controller knows about every other
  controller. Twenty repetitions per scenario.}
\label{fig:interControllerDiscovery}
\end{figure}

Fig.~\ref{fig:interControllerDiscovery} shows the distributions.
Patched-build medians are $4.2$~s (2c\_Simple), $5.9$~s (4c\_Mesh), and
$3.6$~s (5c\_Large); stock-build medians are $10.2$~s, $11.8$~s, and
$8.0$~s, $2.0$--$2.4\times$ slower because the cascade-bound bootstrap of
Section~\ref{sec:bootstrapScaling} delays the first C-Adv from reaching a
peer's domain. The 5c\_Large topology discovers faster than 4c\_Mesh
despite having more controllers, because its richer connectivity offers
more paths for C-Adv propagation. Discovery cost depends on the
inter-domain diameter, not on the number of alternative paths $K$: each
controller needs only one C-Adv per peer, and the number of C-Adv
messages per beacon period stays constant per managed switch. Each
C-Adv's payload does grow with the number of controllers---it carries one
entry, a controller and its path vector, per known controller---but for
the handful of controllers these deployments use, this list fits
comfortably within a single packet, so the added overhead is negligible.

\subsection{Controller failure recovery}
\label{sec:controllerFailureRecovery}

When a controller fails, the switches it managed are momentarily
orphaned and must rejoin the control plane through a surviving
controller. The controller-failure recovery procedure of
Section~\ref{sec:cfrProcedure} keeps this outage short: a switch whose
flow entries expire reverts to its preconfigured initial rules, learns
its new upstream port from the next C-Adv that arrives, and re-runs the
bootstrap procedure against a surviving controller. We quantify how long
that sequence takes end to end. Data-plane recovery from link and switch
failures under a single controller is evaluated
in~\cite{periplus-single}; here the failure is the controller itself.

\begin{figure}[h]
    \centering
    \includegraphics[width=0.35\linewidth]{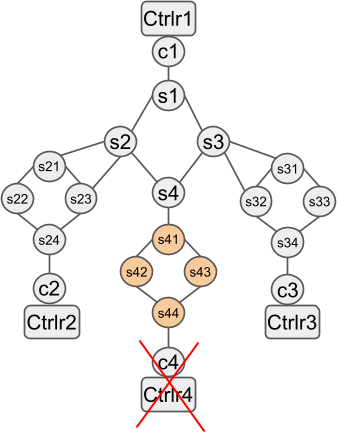}
\caption{4c\_Mesh topology where the controller colocated in c4
	is killed.}
\label{fig:mesh-c4-down}
\end{figure}

The experiment uses the 4c\_Mesh topology. The controller colocated
in c4 is killed, and the four switches it managed (s41--s44) must
reattach via C-Adv to a surviving controller (colocated in c1, c2, or
c3), see Fig.~\ref{fig:mesh-c4-down}.

Because s42, s43, and s44 reach a surviving controller only
through s41 or each other, recovery proceeds as a daisy chain in which
each switch completes its own OpenFlow handshake only after its upstream
neighbour on the new path is itself managed. The metric is the
wall-clock recovery time: the interval between the kill of the controller
colocated in c4 and the
moment the last affected switch is marked managed on its new controller.
We use twenty repetitions per OVS variant.

\begin{figure}[h]
    \centering
    \includegraphics[width=0.65\linewidth]{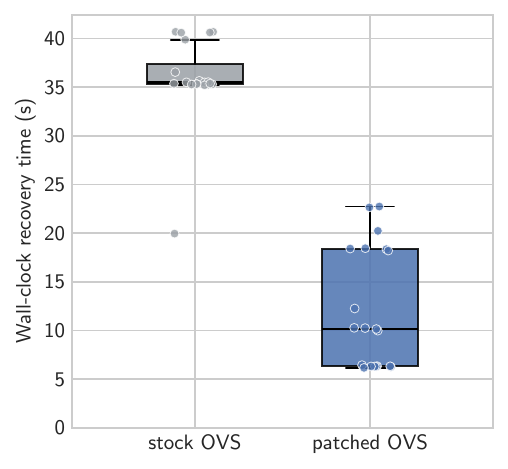}
\caption{Wall-clock controller-failure recovery time after killing the
  controller colocated in c4 in the 4c\_Mesh topology, for the default and
  patched OVS builds. 20
  repetitions per variant.}
\label{fig:controllerFailureRecovery}
\end{figure}

Fig.~\ref{fig:controllerFailureRecovery} shows a median recovery of
$10.2$~s on the patched OVS build: this is the Periplus reattach cost,
the time to install the flows for the four-switch daisy chain plus the
BFD and C-Adv detection latency. The default build is roughly three
times slower (median $35.5$~s), but the gap is entirely OVS-side. Stock
OVS keeps a switch convinced its dead controller is still alive for
several seconds, through hardcoded floors on its inactivity-probe and
reconnect-backoff timers and, dominant here, a definition of connection
``activity'' that counts queued-but-unacknowledged messages as peer
liveness. The optional rconn patch, described in~\cite{periplus-single},
removes these obstacles.

\subsection{Flow-table occupancy by switch role}
\label{sec:flowtableoccupancy}

The amount of per-switch forwarding state a controller must install
bounds how large a network can grow before a single switch's flow table
becomes the limit. Periplus's central claim is that this state is local:
an interior switch holds only the forwarding graph toward its own
controller, while a border switch additionally holds one transit graph
per peer controller it relays for. This is the property that makes the
bootstrap scaling of Section~\ref{sec:bootstrapScaling} possible: because
coordination state is confined to domain boundaries, a network can be
partitioned across more controllers without inflating the state on any
interior switch. We test the claim by counting the distinct $(table,
priority, match)$ rules the controller installs on each switch, grouped
by role: \emph{root} (the controller's co-located switch), \emph{border}
(a switch that received at least one transit-graph install), and
\emph{interior}. We sweep the Mesh family from one to four controllers,
with five repetitions per configuration.

\begin{figure}[h]
    \centering
    \includegraphics[width=0.7\linewidth]{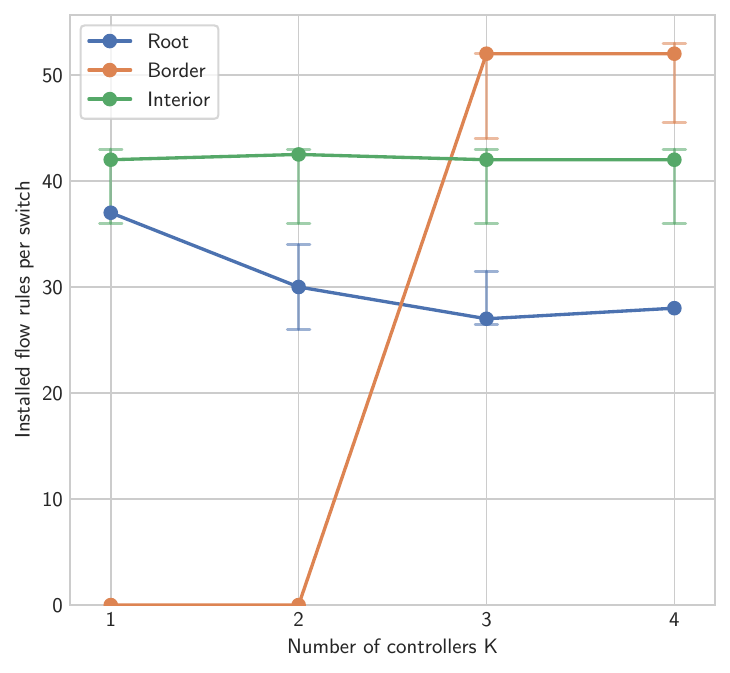}
\caption{Per-switch flow-table occupancy by role as the Mesh family is
  partitioned across one to four controllers. Markers are medians,
  whiskers the inter-quartile range.}
\label{fig:flowTableOccupancy}
\end{figure}

Fig.~\ref{fig:flowTableOccupancy} confirms the locality claim. Interior
occupancy is constant at $\approx$42 rules per switch regardless of the
number of controllers. 

Border switches sit at $\approx$52 rules, the
interior baseline plus a single peer's transit subgraph, and appear only
once a domain must relay for a non-adjacent peer (with two controllers
placed at opposite ends of the Mesh, no switch lands on a transit path,
so border switches first appear at three controllers). The root switch
holds $\approx$27--37 rules, dropping slightly as controllers are added
and routing work is shared among peers. None of the three counts depends
on the absolute number of switches in the network: per-switch state is
set by a switch's role, not by network size, which is precisely what
lets Periplus partition a network across controllers and scale to
networks of around a hundred switches. By contrast, the
spanning-tree-per-controller designs of Section~\ref{sec:related-work}
install one tree on every switch per controller, so their interior state
grows linearly with the number of controllers; Periplus keeps it
constant.

\section{Conclusions and Future Work}
\label{sec:conclusions}

This paper presented the multi-controller coordination plane of
Periplus, an in-band SDN control plane whose single-controller design is
developed in~\cite{periplus-single}. Periplus partitions a wide-area
network across several controllers while keeping the coordination cost
in-band and local: controllers discover one another through Controller
Advertisement messages, and inter-controller routes are assembled
incrementally from partial forwarding graphs inserted by border
switches, so that per-controller forwarding state is confined to the
switches at domain boundaries and never distributed across the interior
of an intermediate domain.

The Mininet evaluation, including a large-scale scenario with 96
switches and up to five controllers, confirms the two properties this
design targets and characterises the coordination mechanism in action. Partitioning scales bootstrap: a 96-switch network is
brought under control in around ten seconds once it is split across
three or more controllers, since each controller bootstraps only its own
domain. Per-switch flow-table occupancy is set by a switch's role rather
than by network size: interior occupancy is constant regardless of the
number of controllers, and grows only at border switches, by one transit
subgraph per peer controller. Inter-controller discovery converges in
3.6--5.9\,s, and a failed controller's switches reattach to a surviving
controller in a median of 10.2\,s. The bootstrap, discovery, and
recovery figures above are all reported on an optional patched OVS build;
the patch addresses stock-OVS reconnect heuristics and is not required by
the Periplus design, which otherwise runs unmodified on stock OVS.

Several directions remain open. The evaluation so far relies on Mininet
emulation; validating Periplus on a physical testbed---in particular over
the resource-constrained links of the rural deployments that
motivate it---is an important next step. The current first-come,
first-served switch assignment leaves load uneven across controllers; using the
switch-count and topology information already exchanged through C-Adv to
inform attachment decisions is a natural step toward load-aware
controller assignment. Building coordination and consistency protocols,
such as eventual-correctness approaches~\cite{panda2017scl}, on top of
Periplus's inter-controller communication infrastructure is a further
open direction.

\bibliographystyle{plain}
\bibliography{refs}

\end{document}